\documentclass[prb,twocolumn,amsmath,amssymb,floatfix,showpacs]{revtex4}

\usepackage{graphicx}

\usepackage{bm}

\usepackage{amssymb}

\usepackage{color}
\usepackage{epsfig}

\begin{document}

\title{A mean-field model for the electron glass dynamics}

\author{Ariel Amir, Yuval Oreg, Yoseph Imry}

\affiliation { Department of Condensed Matter Physics, Weizmann
Institute of Science, Rehovot, 76100, Israel\\}

\newcommand{\be}{\begin{equation}}
\newcommand{\ee}{\end{equation}}

\begin{abstract}
We study a microscopic mean-field model for the \emph{dynamics} of
the electron glass, near a local equilibrium state. Phonon-induced
tunneling processes are responsible for generating transitions
between localized electronic sites, which eventually lead to the
thermalization of the system. We find that the decay of an excited
state to a locally stable state is far from being exponential in
time, and does not have a characteristic time scale. Working in a
mean-field approximation, we write rate equations for the average
occupation numbers $\langle n_i \rangle$, and describe the return to
the locally stable state using the eigenvalues of a rate matrix,
$A$, describing the linearized time-evolution of the occupation
numbers. Analyzing the probability distribution $P(\lambda)$ of the
eigenvalues of $A$ we find that, under certain physically reasonable
assumptions, it takes the form $P(\lambda) \sim
\frac{1}{|\lambda|}$, leading naturally to a logarithmic decay in
time. While our derivation of the matrix $A$ is specific for the
chosen model, we expect that other glassy systems, with different
microscopic characteristics, will be described by random rate
matrices belonging to the same universality class of $A$. Namely,
the rate matrix has elements with a very broad distribution, i.e.,
exponentials of a variable with nearly uniform distribution.

\end{abstract}

%Amorphous semiconductors, metallic glasses, glasses
%Electronic transport phenomena in thin films
%Mobility edges; hopping transport

 \maketitle

 \section{Introduction}

Experiments conducted on thin-films of amorphous or crystalline
semiconductors such as indium-oxide or silicon, show that when
driven out-of-equilibrium (for example by shining light on the
system or by changing a gate voltage), the system exhibits slow
relaxations, observable on the scale of minutes or hours \cite
{{monroe1}, {Zvi:1}}. In many cases a logarithmic or weak power-law
time-dependence of the measured quantity (such as conductance and
capacitance) is observed over many decades of time
\cite{{Martinez},{Zvi:5},{silicon_glass}}. A common feature of the
experimental systems is that they are highly disordered, so that
most electronic states are localized. If the carrier concentration
is high enough \cite{condition_of_high_density}, the (unscreened)
Coulomb interactions may play an important role \cite{Zvi:2}. This
system is usually referred to as the electron glass, since it
exhibits many features characteristic of glassy systems: memory
effects~ \cite{Zvi:4} (the conductance depends on the previous
perturbations applied to the system) and aging \cite{Zvi:3} (the
duration of time of the perturbation is applied affects the
relaxation timescales). Similar effects have been observed in
granular Al \cite{{grenet1},{grenet2}} showing that the underlying
principles may be more general.

In this paper we study a mean-field model for the dynamics of the
system. A variety of systems in nature can be described, near a
locally stable state, by a matrix equation of the type: \be \frac{d
\vec{\delta n}}{dt}= A \cdot \vec{\delta n} \label {matrix_eq} ,\ee
the component $\delta n_i=n_i - f_i$ is the deviation of the average
occupation of the i'th site, $n_i$, from its value $f_i$ at the
locally stable point. The local stability of the point implies that
the matrix $A$ must have only non-positive eigenvalues, and, for
large systems, their distribution will determine the average
time-dependence of the return to the locally stable point, after the
system was slightly pushed away from it. It must be emphasized that
our approach is different from the usual theoretical explanations of
aging phenomena in glasses, in which the system explores the energy
landscape, and slow relaxations are a result of the existence of
many metastable states. In our model, the system is found in the
vicinity of one locally stable point, at all times (we do not use
the term metastable to stress this difference). This assumes that
the initial perturbation is small enough (and so is the
temperature), such the system does not reach other (lower) minima,
but remains in the same region of phase space. Slow relaxations are
due to isolated states that, statistically, happen to have a long
life time. It should be emphasized that although the interactions
lead to the non-trivial Coulomb gap \cite {efros} in the equilibrium
state, the slow dynamics will occur also without interactions.

 If it is given that the distribution of eigenvalues diverges at small (negative)
eigenvalues, and is of the form $P(\lambda) \sim
\frac{1}{|\lambda|}$ (as happens in our model) it is straightforward
to see that logarithmic relaxation in time, in an appropriate time
window, is obtained (assuming the eigenvectors are excited with
uniform probability). In this work we show that starting from a
realistic microscopic model for the electron glass system, the
described situation indeed occurs, and we argue that it is plausible
that other physical systems will also show similar results.

The structure of the paper is as follows. The model is defined in
\ref {definition}. In \ref {mean-field-eq}, we review the
application of the mean-field approximation to the peculiar
equilibrium properties of the system, manifesting the Coulomb gap.
In \ref {mean-field-steady}, we briefly discuss the mean-field
steady-state solution in the presence of an external field, leading
to the Miller-Abrahams  model \cite {miller_abrahams}.

In a similar fashion, in section \ref{dynamics} we suggest to study
the dynamics of the system by writing a set of ordinary differential
equations, described by Eqs. (\ref{rates}) and (\ref{decay}), giving
the time-evolution of the occupancies of the localized states. This
is already an approximation neglecting interference or quantum
fluctuation effects. In \ref{application_to_electron_glass}, we
study the dynamics of the electron glass, starting from an
out-of-equilibrium state. Linearizing Eq.  (\ref{rates}), we obtain
the time-evolution equations of the occupations and obtain Eq. (\ref
{matrix_eq}), with the random matrix $A$ belonging to a different
class from the usual gaussian random matrix ensembles. The
statistics of the eigenvalues is studied numerically, see Fig.
\ref{distribution}. In \ref{eigenvalues} we study a simplifying
limit, analytically. Both lead to a distribution of eigenvalues
$P(\lambda)$ diverging at low values (down to a cutoff), leading to
slow relaxations of the physical observables, as seen
experimentally. This behaviour might be characteristic of glassy
systems. Finally, in \ref {conductance}, we discuss the relation
between the relaxation of the occupation numbers and the
conductance.

\section {Mean-field model for electron glass} \label
{mean-field-model} In this section we discuss a specific microscopic
model for the dynamics of the electron glass. We will show that it
leads to a rate equation of the type of Eq.  (\ref {matrix_eq}), and
find explicitly the matrix elements.
 \subsection{Definition of
the model} \label {definition} We study a system of $N$ localized
states and $M < N$ electrons, with a coupling between the electrons
and a phonon reservoir. Since the states are localized, the
electrons will interact via an unscreened Coulomb potential. In the
absence of electron-electron interactions, the localized states have
different energies, $\epsilon_i$, due to the disorder. Our model
also contains structural disorder: the positions of the sites are
assumed to be random. Although localized states are orthogonal,
their tails overlap, and therefore phonons may induce transitions
between them. The generic coupling between electrons and phonons is
given by the form $\sum_q M_q c_i^{\dag}c_j (b^{\dag}_q+b_{-q})$
where $c_i^{\dag},c_j$ are electron creation and annihilation
operators at local sites $i,j$ and $b_q$ annihilates a phonon. $M_q$
is a coefficient accounting for the strength of the electron-phonon
coupling.

Let us denote the energy difference of the electronic system before
and after the tunneling event by $\Delta E$, containing the
interaction effects. For weak electron-phonon coupling ($|M_q|^2 \nu
\ll \Delta E$, where
 $\nu$ is the phonon density
of states), the transition rate $\gamma_{ij}$ of an electron from
site $i$ with energy $E_i$ to site $j$ with energy $E_j<E_i$ a
distance $r_{ij}$ away, can be calculated treating the coupling as a
perturbation. This yields, up to polynomial corrections \cite
{efros}:

\be \gamma_{ij} \sim |M_q|^2 \nu f_i (1-f_j)
e^{-\frac{r_{ij}}{\xi}}[1+N(\Delta E)], \label {rates} \ee where
$f_i$ is the Fermi-Dirac distribution. For upward transitions
($E_j>E_i$) the square brackets are replaced by $N(\Delta E)$. These
rates may be renormalized due to polaron-type orthogonality effects
\cite{leggett_review}.

We will be interested in the dynamics of the system when it is
out-of-equilibrium, namely in the time-dependence of the occupation
numbers and the conductance after an initial excitation. But we
first show how non-trivial equilibrium properties are obtained from
the mean-field picture.

 \subsection{Equilibrium properties near a locally stable point}
 \label {mean-field-eq}

 In an approximation similar to those used in spin glass theory \cite{spin_glass},
 we define $f_i= \langle n_i \rangle$,  where $n_i$ the site occupation (which takes the values zero or one) and
$\langle \rangle$ denotes averaging over a time scale much larger
than that related to the frequency of the phonon processes but
smaller than the relevant scale for the observation of the dynamics.
This is a mean-field type approximation, and may be used regardless
of the interactions in the system.  Let us first discuss the thermal
equilibrium state, near the locally stable point. The sites
occupation must follow the Fermi-Dirac distribution and therefore:

\be f_i (E_i)= \frac{1}{1+e^{\frac{E_i-\mu}{T}}}, \label {fermi} \ee
where $\mu$ is the chemical potential, and the Boltzmann constant is
set to be one.

In the mean-field approximation we can calculate the average
potential energy of site $i$:

\be E_i= \epsilon_i + \sum_{j \neq i} \frac{e^2 f_j}{r_{ij}} \label
{energies}.\ee

This approximation improves as the number of interacting sites
increases \cite{mean_field_infinite}. The long range nature of the
interaction means that the energy of a site will be determined by
many of its neighbours, and gives intuitive justification of the use
of mean-field theory. Nevertheless, we should emphasize that this is
an uncontrolled approximation, and the limits of its validity should
be checked.

Combining Eq. (\ref {fermi}) and  (\ref {energies}), one obtains a
self-consistent equation for the energies. It is common to use an
unbiased disorder distribution, and add a background charge $\nu$ to
each site \cite {coulomb_gap_mean_field}. In the mean-field picture,
this will lead to the equation:

\be E_i = \epsilon_i + \sum_j \left(
\frac{1}{1+e^{\frac{E_j-\mu}{T}}}\ -\nu \right) \frac{e^2}{r_{ij}}
 \label
{solve_E}. \ee For half filling, $\mu=0$, $\nu=0.5$.

 One should
notice that there are many solutions to this self-consistent
equation. Rigorously, one cannot call any solution an equilibrium
distribution, since the equilibrium distribution is a Boltzmann
average over \emph{all} configurations, not only those near the
locally stable point. The solution may be viewed as a 'local
equilibrium'. We will see that the physical picture obtained is
quite plausible. At low temperatures \cite{T_c}, the probability
distribution of the energies will contain a soft gap at the Fermi
energy, known as the Coulomb gap \cite {{efros},{efros_sce},
{coulomb_gap_mean_field},{peierls},{pankov},{malik},{coulomb_gap_experimental},{coulomb_gap_exp2}}.

\begin{figure}[b!]
\includegraphics[width=0.5\textwidth]{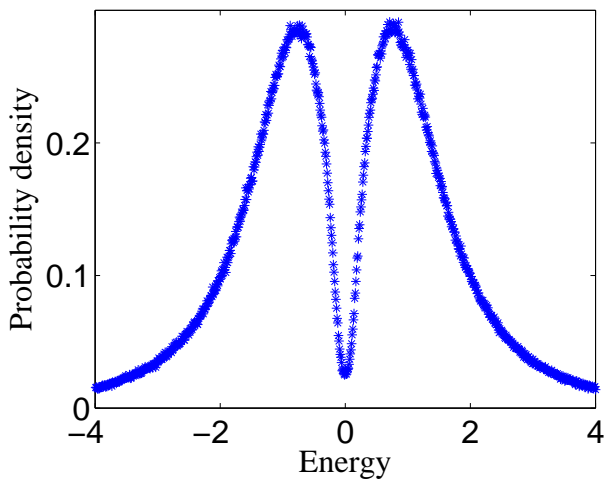}
\caption{Histogram of the site energies in two dimensions, for
N=10000, $\mu=0$ (half filling), obtained by solving the
self-consistent Eq. (\ref {solve_E}). The sites were uniformly
distributed in a square, with $\frac{e^2}{r_{nn} T}=20$, where
$r_{nn}$ denotes the average nearest-neighbor distance and $T$ is
the temperature. The energies $\epsilon_i$ were uniformly
distributed in the interval $[-\frac{W}{2},\frac{W}{2}]$, with
$W=1$. The y axis denotes the probability density of the energies
$E_i$. The graph is the average over 300 instances. Notice the
finite value of the density at the minimum, due to the finite
temperature. }\label{coulomb_gap_2d}
\end{figure}

Eq. (\ref{solve_E}) can be solved numerically by starting with a
random set of energies, and evolving them iteratively, within the
mean-field model. This was done following Ref. [\onlinecite
{coulomb_gap_mean_field}], by solving the equations for many random
instances, and averaging over them. In this way a histogram of the
on-site energies is obtained. When normalized correctly, it gives
the single-particle DOS (density of states), as function of energy.
The results for two-dimensions, yielding the Coulomb gap, are given
in Fig. \ref{coulomb_gap_2d}. Notice that the obtained DOS contains
a linear gap near the Fermi energy, in accordance with other works
\cite {{efros_sce}, {coulomb_gap_experimental}}.

 \subsection{Response to an external field: steady-state solution}  \label {mean-field-steady}

  When a small electric field is applied,
 there are corrections to the average occupations and also to the average
energies. It can be shown that the problem of finding the
steady-state solution corresponds to that of solving the
steady-state of a resistance network, using Kirchoff's laws
\cite{miller_abrahams}. The solution, when neglecting interactions,
gives the well-known Mott Variable Range Hopping \cite {mott}, which
was observed experimentally in many cases \cite {mott_exp}. We
emphasize that this calculation is in fact a mean-field one: the
steady-state solutions are obtained from time-dependent equations
which are essentially the mean-field equations. In the following we
propose to use the same ideas to discuss the dynamics of the system
out-of-equilibrium.

\section{The Dynamics} \label {dynamics}

Let us pose the following question: How will the occupation numbers
or conductance depend on time, when the system is pushed slightly
out of the locally stable point? Having seen that the mean-field
approximation yields the correct density of states as well as the
out-of-equilibrium steady-state solution in the presence of an
external field, we propose to use the same approximation to describe
the dynamics of the systems, when prepared out of local equilibrium.

Experimentally, the form of the relaxation depends on the details
and mechanism of the excitation. For simplicity, let us assume that
the initial perturbation takes the form of a random addition $\delta
n_i$ to the state occupations, with $\sum_i \delta n_i =0$,
reflecting particle number conservation.

Assuming that the initial change in the occupations is small, we can
still use Eq. (\ref{rates}) for the tunneling rates, with the
average occupations at the locally stable point $f_j$ substituted by
the occupation numbers slightly out of equilibrium $n_j$ (which can
take any value between 0 and 1). The energies at each instance are
related to the out-of-equilibrium occupations by Eq.
(\ref{energies}), upon replacing $f_j$ by $n_j$, and we can write
the time-evolution of the average occupation as:

\be \frac{d n_i}{dt}= \sum_{j\neq i}\gamma_{ji}-\gamma_{ij} . \label
{decay} \ee

This defines the problem completely. At the locally stable point
itself, the RHS of Eq. (\ref {decay}) must vanish. Therefore not too
far from the point, we can take the first (linear) order in the
quantities $\delta n_i$, the deviations from the stable point. The
linearized equation then takes the form of Eq. (\ref {matrix_eq}),
where $\vec{\delta n}$ is a vector of the deviations of the
occupation numbers from their local equilibrium values. The $N$
eigenvalues and eigenvectors of the matrix $A$ will determine the
decay rates of the system.

We would like to stress that the matrix $A$ can be calculated for
the cases of interest, by linearizing the equations of motion
\emph{near the locally stable point}. This strategy is completely
general, and will be valid for any system which can be described by
equations of motion, and has a locally stable point (such would be
the case for most classical systems, and many quantum systems in a
mean-field approximation). The dynamics of the system when pushed
slightly away from the fixed point, will be characterized by the
eigenvalues of the rate matrix. In the common case where disorder
plays a role, the dynamics will depend on the distribution of
eigenvalues of the matrix, thus, we obtain a problem of random
matrix theory \cite {mehta}, where the eigenvalue distribution is
responsible for the \emph{dynamics} of the system. An extremely
relevant property of the electron glass case, as we shall
demonstrate in \ref {application_to_electron_glass}, is that the
entries of the random matrix are \emph{exponentials} of the broadly
distributed parameters (energy and distance). Another important
feature of these matrices is that the sum of every column vanishes.
These properties make this matrix belong to a different class from
the usual gaussian ensembles treated in random matrix theory, and
will play an important role in the dynamics, leading to slow
relaxations. A similar class of matrices was studied previously by
Mezard et al. \cite {distance_matrices}.

In the following section we will derive the form of the matrix $A$
for the particular case of localized states coupled due to phonons.
We will find that the probability distribution is divergent for
small eigenvalues, and suggest what the minimal properties leading
to such a distribution are. The implications of this distribution on
the time-dependent relaxation of the occupation numbers and
conductance will then be discussed.

\subsection {Application to the electron glass model} \label
{application_to_electron_glass}

Starting from Eq. (\ref {rates}), a calculation of the elements of
matrix $A$ in Eq. (\ref{decay}) shows that:

\be A_{ii}= \sum_{j \neq i} -\frac{\gamma^0_{ij}}{n_j (1-n_j)} %+ \frac{\Gamma_{ij}}{kT} \frac{e^2}{r_{ij}}
 ,\ee  where $\gamma^0_{ij}$ are
the local equilibrium rates, given by Eq. (\ref{rates}).  For $i
\neq j$:

\be { A_{ij}= \gamma^0_{ij}\frac{1}{n_j (1-n_j)} -\sum_{k \neq j, i}
\frac{e^2 \gamma^0_{ik}}{T}( \frac{1}{r_{ij}}- \frac{1}{r_{jk}})}.
\label{realistic} \ee

Notice that the matrix is not symmetric, due to the $n_j (1-n_j)$
term. The sum of each column of the matrix vanishes, guaranteeing
particle number conservation.

\begin{figure}[b!]
\includegraphics[width=0.5\textwidth]{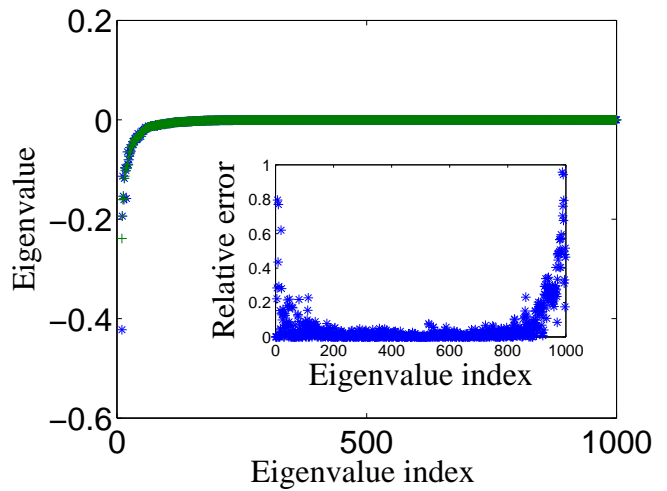}
\caption{Comparison of the eigenvalues of the full linearized matrix
(stars) and the ones obtained after neglecting the 'Coulomb' term
(crosses), the second one in Eq. (\ref{realistic}). The difference
is mostly seen for the large magnitude eigenvalues.  $N=1000$,
$\frac{\xi}{r_{nn}}=0.1$ and $\frac{e^2}{r_{nn} T}=10$. The energies
$\epsilon_i$ were uniformly distributed in the interval
$[-\frac{W}{2},\frac{W}{2}]$, with $\frac{W}{T}=10$. The inset shows
the relative error in replacing the full matrix with the
approximated one, defined as the difference between the
approximation and exact diagonalization, divided by the exact value.
}\label{coulomb_important}
\end{figure}

\emph{A-priori} one would expect that at low enough temperatures $T
\ll \frac{e^2}{r_{nn}}$, we could neglect the first term in the
equation for the regime of interest. However, at low temperatures
the occupations of the sites tend to 0 or 1 exponentially, meaning
the $\frac{1}{n_j(1-n_j)}$ term explodes much faster than the
$\frac{1}{T}$ part in the second term. Viewed in a different way, if
one looks at the expression of the mean-field rates $\gamma_{ij}
\sim n_i (1-n_j)[1+N(\Delta E)]$, one sees that if two states are
close in energy, then the first term in the matrix element
(\ref{realistic}) $\sim \frac{n_i}{n_j}N(\Delta E)\sim
\frac{T}{\Delta E}$. Therefore there is good coupling between
\emph{any} two states close in energy (and distance), not only those
ones close to the Fermi level, as is the case for the second term.
Therefore the 'phase space' is much larger for the first term, and
the second one can be neglected. We have calculated numerically the
eigenvalue distribution for some specific system parameters, and
indeed it was found that the second part has a small influence on
most eigenvalues, see Fig. \ref {coulomb_important}.

An important property follows: the off-diagonal elements are
positive. Together with the property that the sum of every column
vanishes, the stability of the mean-field solution is guaranteed:
all the eigenvalues are negative, characterizing decay. A proof of
the statement can be found in Ref. [\onlinecite
{distance_matrices}].

\begin{figure}[b!]
\includegraphics[width=0.5\textwidth]{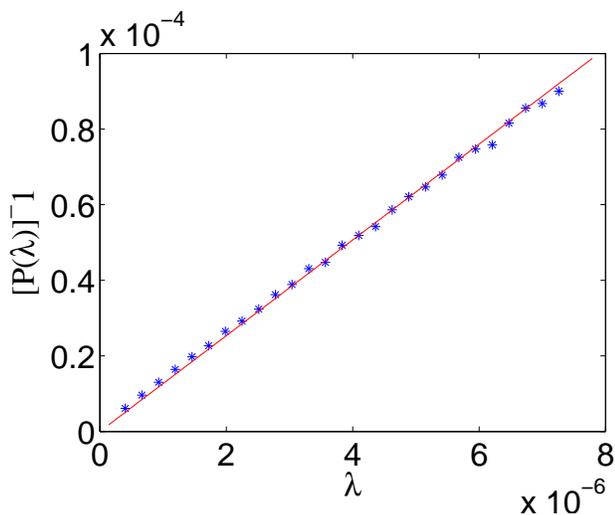}
\caption{Distribution of eigenvalues obtained numerically, averaged
over 1000 realizations. The parameters were as for Fig.
\ref{coulomb_important}. The fit to the reciprocal distribution is
linear.} \label{distribution}
\end{figure}

Let us consider the distribution of the eigenvalues. Fig. \ref
{distribution} shows the distribution of eigenvalues of the matrix
$A$, as obtained numerically. We first found a mean-field solution
by iterating the equations (see, for example, \cite
{coulomb_gap_mean_field}), then used Eq. (\ref {realistic}) to
construct the relevant matrix. The eigenvalues of this matrix were
then found numerically. Notice that the localization length $\xi$
influences the dynamics, although it has no effect on the
equilibrium properties, at least as long as the localized states are
spatially well separated.

In \ref{eigenvalues} we will analyze a simplifying limit, when the
rather complicated dependence on energy can be neglected, and
consider only the exponential dependence of the tunneling rate on
length. Both limits give approximately a $\frac{1}{|\lambda|}$
distribution (up to logarithmic corrections), reminiscent of
$\frac{1}{f}$ noise \cite{1_f}. This suggests that the result may be
more general, and not dependent on the details of the specific
model. Note that the interactions affect the mean-field solution
(and the Coulomb gap), but the calculation shows the slow dynamics
will exist also without them.

Let us now discuss the consequences of this distribution for the
dynamics of the system. Having found that the distribution is $\sim
\frac{1}{|\lambda|}$ down to some minimal value, we conclude that if
all eigenvectors (except the one with eigenvalue zero, not
conserving particle number), are excited with equal probability, the
time-evolution of the deviation from the local stable point will be
related to the Laplace transform of the eigenvalue distribution.
This will give rise to logarithmic decays. Let us show this in more
detail: if the eigenvectors are denoted by $\vec{n_\lambda}$, the
time-dependent deviation from the locally stable point will be given
by:

\be \vec{\delta n} = \sum_\lambda c_{\lambda} \vec{n_\lambda} e^{-
\lambda t}. \ee

We shall assume the eigenvectors are excited with roughly uniform
probability.  Going to the continuous limit, and utilizing the fact
that the components of the eigenvectors themselves are also random
variables, we obtain that the norm of $\vec{ \delta n}$ should relax
as:

\be  |\vec{ \delta n}(t)| \sim \int_{\lambda_{min}}^{\lambda_{max}}
\frac{e^{- \lambda t} }{\lambda} d \lambda \sim \gamma_E-\log [ t
\lambda_{min}], \ee for $\frac{1}{\lambda_{min}}<t <
\frac{1}{\lambda_{max}}$.

\subsection {Dynamics of exponential models} \label {eigenvalues}

Hitherto we have discussed a specific model for glass dynamics in a
system constructed from interacting electrons and phonons. The
actual form of the rate matrix eigenvalue distribution did not
depend strongly on the details of the matrix elements. In this
section we will show that there are few sufficient conditions on the
random rate matrix $A$ that will make the relaxation process long.

Let us look at the dynamics which follow from a class of random
matrices obeying the following properties:

1. The sum of every column vanishes. This follows from particle
number conservation.

2. The entries of the matrices are distributed over a very broad
range. This happens, for example, when they are exponentials of a
more or less flat distribution \cite{pollak}.

We expect that a variety of systems that exhibit a glassy behaviour
may be described by a random rate matrix belonging to this class.
The matrix obtained for the electron glass system indeed obeys these
properties. The first property was shown explicitly in section
\ref{application_to_electron_glass}. To see the second, let us
examine Eq. (\ref{realistic}). If we neglect cases where the
energies $E_i$, $E_j$ and $|E_i-E_j|$ are smaller than $T$, we can
recast the equation into a more transparent form:

\be A_{ij} \sim
e^{-\frac{r_{ij}}{\xi}}e^{\frac{-|E_i-E_j|-|E_i|+|E_j|}{2T}}. \label
{simplified}\ee

Due to the exponential, the matrix entries are indeed broadly
distributed.

We shall now discuss a specific class of matrices which can be
analyzed analytically. As seen in Eq. (\ref{simplified}), the matrix
elements for the electron glass system contain a factor
$e^{\frac{-r_{ij}}{\xi}}$. If $\xi$ is much smaller than the typical
distance $r_{nn}$, it is plausible that this factor would be
dominant in determining the eigenvalue distribution. This motivates
us to discuss a simpler model, of so-called distance matrices
\cite{distance_matrices}: assume we have $N$ random points in a two
dimensional space. Let us define a matrix $B_{ij}=
e^{-\frac{r_{ij}}{\xi}}$, where $r_{ij}$ is the distance between
points $i$ and $j$, and $\xi$ some constant. Let us choose the
diagonal elements of the matrix such that the sum of every column
vanishes. Following the previous discussion of the dynamics, we are
interested in the distribution of eigenvalues of such a matrix. A
mapping of this problem to a field theory problem is given in Ref.
[\onlinecite {distance_matrices}], enabling one to look at a
low-density approximation to the theory. Mezard et. al calculate the
resolvent $R= \frac{1}{N} Tr \frac{1}{\lambda-H}$, the imaginary
part of which yields the density of states, i.e, the distribution of
eigenvalues. Using their formula (21) for the case of $f(r)=
e^{-\frac{r}{\xi}}$, we obtain that the low-density expansion of the
resolvent is:

\be R(\lambda)= \frac{\rho}{2 V}\int  dx dy \left(
\frac{1}{\lambda+2 e^{-\frac{r}{\xi}}}-\frac{1}{\lambda} \right) ,
\label{parisi_formula}\ee  where the integrals are performed over
the whole volume. The second term gives rise to a delta function at
the origin, which comes from the zero eigenvalue the matrix always
possesses, and is of no particular interest since the eigenvector
associated with this eigenvalue cannot be excited while preserving
the particle number. The condition for the approximation to be valid
is $\xi \ll r_{nn}$, as we shall show later in a more transparent
way.

Since the density of states is given by
$-\frac{Im[R(\lambda)]}{\pi}$, we can use the fact that
$Im[\frac{1}{x+i \epsilon}]= -i \pi \delta(x)$ and obtain the DOS
as:

\be P(\lambda)= \frac{\rho}{2 V}\int  dx dy \delta(\lambda+2
e^{-\frac{r}{\xi}}). \ee %%

Performing the integral in one-dimension, for eigenvalues not too
close to the minimal value $2e^{-\frac{L}{\xi}}$, leads to the
result: \be P(\lambda)=\frac{-N \xi}{L \lambda}  , \label {1d} \ee
with $\lambda$ in the interval $[-2, -2 e^{\frac{-L}{\xi}}]$.

Repeating the calculation in two-dimensions, again, for eigenvalues
not too close to the minimal values, yields: \be P(\lambda)=
\frac{\pi N \xi^2 \log(-\frac{\lambda}{2})}{L^2 \lambda}. \label
{2d} \ee

We shall now give a transparent demonstration of these results. In
the low density limit, we can couple each site to its
nearest-neighbor, thus dividing the system into $\frac{N}{2}$ pairs.
Neglecting the effect of all other sites, we obtain that the
eigenvalues will be similar to those of an ensemble of 2X2 matrices
of the form:

\be M=\left(
\begin{array}{cccc}
  -e^{\frac{-|x-y|}{\xi}} & e^{\frac{-|x-y|}{\xi}} \\
  e^{\frac{-|x-y|}{\xi}}& -e^{\frac{-|x-y|}{\xi}} \\
\end{array}%
\right).\ee

Since the matrix elements of $B$ decay on the scale of $\xi$, this
would be a good approximation for $\xi \ll r_{nn}$. One eigenvalue
of $M$ is 0, and the other is $-2e^{\frac{-|x-y|}{\xi}}$. Therefore
half the eigenvalues of $B$ will be the zero under this
approximation, and the distribution of the other eigenvalues will be
that of the random variable $-2e^{\frac{-r}{\xi}}$, where $r$ is the
nearest-neighbour distance. Notice the zero eigenvalues correspond
to the second term in Eq. (\ref{parisi_formula}).

The distribution of the nearest-neighbour distance can be
calculated: looking at a typical site, let us calculate the
probability that its nearest-neighbour is at least distance $r$
away. This is equivalent to asking that \emph{all} of its neighbours
are at least a distance $r$ away, and since they are randomly
distributed, we obtain that:

\be Prob(r)= (1- V_D \frac{r^D}{L^D})^{N-1},\ee where $V_1=2$ and
$V_2= \pi$.

For $r \ll L$ we can approximate:

\be Prob(r)= e^{- V_D N \frac{r^D}{L^D}}. \label {prob}\ee

We have assumed that the initial site is not too close to the
boundaries. Since we are interested in the probability, the sites
near the boundary will give a negligible correction to the above
probability: the sites for which Eq. (\ref{prob}) fails are a
distance of order $r_{nn}$ or less from the boundary. Therefore
their fraction in the system is of order $\frac{r_{nn}}{L} $. For $N
\gg 1$, this fraction is negligible.

The probability \emph{distribution} can be calculated by
differentiating with respect to $r$, leading to:

\be P(r)= V_d D N \frac{r^{D-1}}{L^D} e^{- V_D N \frac{r^D}{L^D}}.
\ee

By construction, the probability distribution is exactly normalized.

In one-dimension, the eigenvalue distribution that follows is:

\be P(\lambda)=-N\frac{e^{\frac{-2N \xi}{L}
|\log(-\frac{\lambda}{2})|}}{L \lambda} \xi \sim
\frac{1}{\lambda^{1-\epsilon}}, \label{1_d_formula} \ee

with $\epsilon=\frac{2 N \xi}{L} \ll 1$,  while for two-dimensions:

\be P(\lambda)= \frac{\pi N \xi^2 \log(-\frac{\lambda}{2}) e^{-\pi
\frac{\xi^2}{L^2} N \log^2(-\frac{\lambda}{2})}}{L^2 \lambda}.
\label {2_d_formula} \ee

\begin{figure}[b!]
\includegraphics[width=0.5\textwidth]{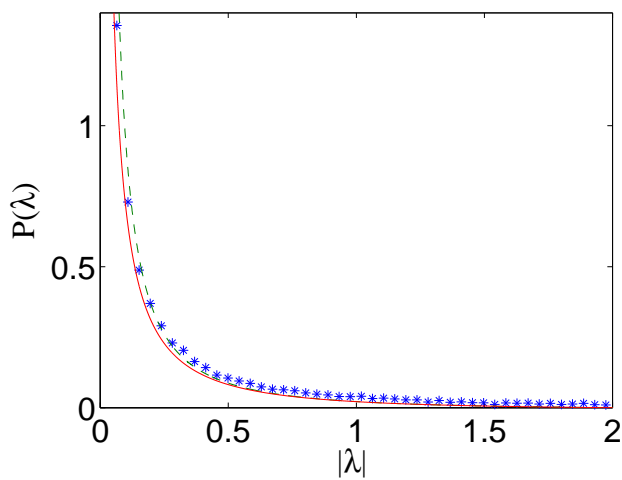}
\caption{Distribution of the eigenvalues in two dimensions, for a
low-density ($\frac{\xi}{r_{nn}}=0.05$). $N=1000$, and the results
were obtained after averaging over 1000 instances. The points were
distributed uniformly in the unit square,
 and the matrix element $A_{ij}=e^{-\frac{d_{ij}}{\xi}}$, with
 $\xi =0.0016$. The dashed curve is a plot of Eq. (\ref{2d}),
%$\frac{\log(\frac{-\lambda}{2})}{\lambda}$ distribution,
while the solid curve is a plot of Eq. (\ref {2_d_formula}).
}\label{low_density}
\end{figure}

Aside from the exponential term, Eqs. (\ref{1_d_formula}) and
(\ref{2_d_formula}) coincide with the field-theory Eqs. (\ref{1d})
and (\ref{2d}). Notice that in the latter there is a cutoff on the
eigenvalue magnitude, while for Eqs. (\ref{1d}) and (\ref{2d}) the
distribution is nonzero also for very small eigenvalues. Fig. \ref
{low_density} shows the results of numerical simulations for the
case of low-density distance matrices in two-dimensions, and a
comparison to the theory.

\subsection{Relaxation of the conductance} \label {conductance}
In many cases a logarithmic relaxation in time is observed for the
conductance \cite {{Martinez},{Zvi:5},{grenet2}}. One should ask
what is the relation between the relaxation of the conductivity and
that of the occupations, for the electron glass model.

We now present an intuitive argument motivating the speculation that
the time-relaxation of the conductance should be similar to that of
the occupations. The essence of the argument is the claim that any
perturbation of the equilibrium configuration will lead to enhanced
conductance: if this is true, it is reasonable that as the typical
deviation of the occupation number relaxes, so does the enhanced
conductivity, until it reaches its equilibrium value. For small
enough deviations, the two will be proportional to each other, as
one can always take the lowest order term in the expansion of the
dependence of the out-of-equilibrium conductivity on the occupation
number deviation.

Let us explain our claim that any perturbation will lead to enhanced
conductivity. This may come about by two physically different
mechanism: first of all, we note that when the system is excited, we
create vacant sites well below the Fermi surface, and add electrons
above it. Electrons will tunnel between these sites, and thus even
at very low temperatures current may flow through the system.

The second mechanism is more subtle, and is related to the Coulomb
gap. Let us look at the Einstein relation \cite {imry_mesoscopics},
$\sigma= e^2 \frac{dn}{d\mu} D$. We do not expect to have any
anomalies in the \emph{thermodynamic} DOS, $\frac{dn}{d\mu}$, but
the diffusion constant $D$ should be much smaller. This is because
the single-particle DOS at the chemical potential vanishes: moving
an electron from a site with energy close to the chemical potential
to another site, will necessitate an energy of order of the width of
the Coulomb gap. Therefore the Coulomb gap significantly lowers the
conductivity. We shall assume that for a finite, large enough,
single-particle DOS at $\mu$, the conductivity increases with the
single-particle DOS at $\mu$ \cite{yu} (for systems close to a local
equilibrium, which exhibit the Coulomb gap). The temperature should
be low enough such that the local equilibrium Coulomb gap would not
be smeared \cite {T_c}. Let us suppose that due to our initial
perturbation of the local equilibrium configuration, we have some
excess (positive or negative) $\delta n_i$ in the occupation number
at site $i$. Assuming these numbers to be random with a standard
deviation $\delta n$, the energy at site $j$ will now have an
additional contribution $\sum_i {\frac{\delta n_i}{r_{ij}}}$. A
finite single-particle DOS at $\mu$ will arise, proportional to
$\delta n$. Both mechanisms show that within this model the
conductance relaxation should have a similar time-dependence to that
of the occupation number relaxation.

\section{Conclusions}
We have studied a finite temperature mean-field model for the
 dynamics of the electron glass system. For a perturbation which
 drives the system not too far from the (mean-field) local equilibrium, we
 mapped the problem onto rate equations with a random relaxation matrix $A$. The matrix $A$ belongs to a class different
 than the gaussian random matrix ensembles. We found that the
 distribution of the eigenvalues is approximately $\sim \frac{1}{|\lambda|}$, and naturally yields a logarithmic decay
of the occupation numbers. This may lead to a logarithmic decay of
 the conductance. Such a logarithmic decay of the conductance is observed
 experimentally in many cases. We emphasize the remark made before that the $\frac{1}{\lambda}$ distribution of decay
eigenvalues should be much more general than for the specific model
considered. It might also hold, for example, in the case of
multi-particle transitions, which are believed to be relevant for
the long time properties of glasses.  Further research is needed to
obtain additional predictions of this  model, such as the
time-dependence of the Coulomb gap, and the
 voltage-dependent conductance in the 'two-dip' experiment \cite{Zvi:4}.

\section*{ACKNOWLEDGMENTS}

\normalsize We thank Zvi Ovadyahu, Assaf Carmi and Phillip Stamp for
useful discussions. This research was supported by a BMBF DIP grant
as well as by ISF grants and the Center of Excellence Program.

\bibliographystyle {prsty}

\begin{thebibliography}{10}
\bibitem{condition_of_high_density}
For two-dimensions and weak disorder, the density should be larger
than
  $\frac{T^2}{e^4}$, see \cite {T_c}.
\bibitem{T_c}
The critical temperature in dimension $D$ can be shown to be of
order
  $\left[\frac{n e^{2D}}{\rm{max}(W,e^2
  n^{\frac{1}{D}})}\right]^{\frac{1}{D-1}}$.
\end{thebibliography}

\end{document}